\documentstyle[preprint,aps,eqsecnum]{revtex}
\begin{document}

\title {Fermi edge singularities in 
X-ray spectra of strongly correlated fermions}
\author 
{D. V. Khveshchenko$^{1,2}$ and P. W. Anderson$^1$ \\
%\address
$^1$ Department of Physics, Princeton University, Princeton, NJ
08544, USA\\
$^2$ NORDITA, Blegdamsvej 17, Copenhagen DK-2100, Denmark}

\maketitle

\begin{abstract}
\noindent
We discuss the problem
of the X-ray absorption in a system of interacting fermions and, 
in particular,  
those features in the X-ray spectra that can be used to discriminate  
between conventional Fermi-liquids and novel "strange metals".
Focusing on the case of purely forward scattering off the core-hole potential, 
we account for the relevant interactions in the conduction band by means of the 
bosonization 
technique. We find that the X-ray Fermi edge singularities 
can still be present, although modified, even if the density of states 
vanishes at the Fermi energy, and
that, in general, the relationship between the two appears to be quite subtle. 
 
\end{abstract}

\pagebreak

Since its foundation in the late fifties, the theory of Fermi liquids
has come a long way exploring the limits of its own applicability.
However, the quest for possible departures from the conventional
Fermi liquid behavior still remains one of the mainstreams of the modern 
condensed matter theory.

It is well known that Fermi liquids can be more easily destroyed in low 
dimensions. 
As a common example, in one spatial dimension (1D) the Fermi liquid theory 
utterly fails 
 even for a seemingly innocuous arbitrary weak short-range repulsion.
The corresponding non-Fermi-liquid (NFL) metallic state is characterized by 
power-law decaying
correlation functions and is referred to as the Luttinger liquid (LL)
which can be thought of as a marginal deformation of the Fermi liquid.

Furthermore, the unscreened Coulomb interactions are known to drive 1D fermions 
into yet another,
distinctly NFL, regime where correlations decay faster than any power law.

It is generally believed
that in higher ($D>1$) spatial dimensions 
a NFL behavior can stem from sufficiently 
long-ranged and/or retarded ("singular") interactions.
However, the necessary criteria remain unknown, which leaves room for a 
NFL regime to occur even 
as a result of non-singular, yet sufficiently strong, repulsive 
interactions.

In theory, a NFL behavior is commonly expected to manifest itself in
single electron spectroscopy and, in particular,
in angular resolved photoemission at low energies of incident
photons. However, even for such a widely recognized candidate
for a "strange metal" as the normal state of the high $T_c$ cuprates, direct 
deduction 
of any ultimate evidence of the NFL behavior (e.g., electron self-energy) 
from the photoemission data has proven to be a tedious task \cite{1}. 

In the present paper,  
we discuss a different kind of features that have already been extensively 
studied in a variety of conventional metals: the Fermi-edge singularities (FES) 
in the high-energy X-ray absorption.
First predicted theoretically for weakly interacting (Fermi-liquid-like) metals 
back in 1967 \cite{2,4}, the FES    
became one of the hallmarks of many-body Fermi systems. However, 
the effect of interactions in the conduction band on the FES  
has not drawn much attention, except for the case of
the 1D LLs studied by a number of authors \cite{3,6}. 

In retrospect, the FES provided the first example of a much more generic 
phenomenon
of the "orthogonality catastrophe" (OC). The latter implies that in the 
presence of a sudden perturbation the ground states of an infinite Fermi system
before and after the perturbation was switched on appear to be strictly 
orthogonal to each other.
In the problem of the X-ray-induced photoemission such a step-like 
time-dependent 
perturbation $V(t)=V\Theta(t)$
corresponds to the attractive potential of 
a deep core level stripped off its electron by an incident X-ray photon.

Following the instantaneous shakeup, the initial ground state $|0>$ of the 
Fermi system 
tends to readjust and evolves into the final one $|V>$ by virtue of 
creating coherent multiple particle-hole pairs. These bosonic excitations     
dominate in the action $S_{oc}(t)$ describing the process of relaxation of the 
initial ground state  and lead to the suppression of the time-dependent 
overlap between the two ground states: $<0|V>\sim \exp(-S_{oc}(t))$  at times  
$t\rightarrow \infty$. 

Being closely related to the Green function of a localized 
core hole described by the operator $d(t)$ \cite{4}, 
the overlap factor controls the shape of the    
photoemission peak corresponding to the absorption of  hard 
X-ray photons that knock core electrons 
out of the system  
$$P(\omega)\propto Im \int_0^\infty dt e^{i\omega t}<T d(t)d^{\dagger}(0)> 
\eqno(1)$$
Here $\omega$ is the photon energy measured from the threshold equal to the sum 
of the
binding energy of the core level and the exit work function for the runaway 
electron.

In the standard Fermi liquid case the action $S_{oc}(t)$ diverges 
logarithmically,
which gives rise to the asymmetrical power-law singularity of the absorption 
peak \cite{2,4}  
$$P(\omega)\propto\Theta(\omega) |\omega|^{-1+2\sum_l \delta^2_l/\pi^2}
\eqno(2)$$
that would have been absent in an insulating state where at zero temperature
the peak remains sharp: $P(\omega)\propto \delta(\omega)$.

The partial phase shifts $\delta_l$ in Eq.(2) 
can be expressed in terms of the angular
momentum harmonics of the Fourier transformed 
core hole potential: $\int d{\bf r}V({\bf r})e^{i{\bf r}({\bf k}-{\bf 
k}^{\prime})}
=\sum_lY_l({\hat 
\Omega}-{\hat \Omega}^{\prime})V_l(k_F)$
(throughout this paper we include all the angular momentum
quantum numbers into the definition of $"l"$)
and the density of states (DOS) in the conduction band $\nu(\omega)={1\over 
\pi}\int 
d{\bf k} Im G({\bf k},\omega)$ 
taken at the Fermi energy $\nu_F=\nu(\omega =0)$:
$$\tan \delta_l=-V_l\nu_F  \eqno(3)$$
At first sight, Eq.(3) seems to imply that DOS must remain finite for the 
divergence (2) to occur.
In order to find a possible caveat in this seemingly unavoidable conclusion
we consider the simplest, short-range isotropic,
core hole potential $V({\bf r})=V\delta({\bf r})$ and, 
following the original work by Nozieres and de Dominicis \cite{5},
introduce a transient Green function ${\cal G}(t,t^{\prime}|V)$  which allows 
the 
OC action to be cast in the form: $S_{oc}(t) =\int^V_0dV'\int^t_0 dt^\prime 
{\cal 
G}(t,t^\prime |V')$.

For free fermions this Green function obeys the equation
$${\cal G}(t, t^\prime |V)=G({\bf 0},t-t^\prime)+V \int^t_0 dt^{\prime\prime}
{\cal G}(t, t^{\prime\prime} |V)G({\bf 0},t^{\prime\prime}-t^\prime)
\eqno(4)$$
which relates it to the time dependence of the 
propagator $G({\bf r},t)$ of the conduction band fermions in the absence of the 
core hole potential.

The above, naive, conclusion would then imply that even with the interaction 
effects taken
into account the local value of the propagator 
$G({\bf 0},t)$ would have to retain its Fermi liquid-like $\propto 1/t$ 
behavior, the 
exponent in
the power-law divergence (2) being determined by the corresponding prefactor.  

Below we comment on the case of fermions interacting via $U({\bf r})\propto 
1/r^{2-\eta}$
pairwise potential which demonstrates that the     
above condition may be somewhat more relaxed than
the stringent requirement for a system to remain a Fermi liquid. However, even 
in this "mildly 
NFL" case Eq.(4)  
ceases to be valid, because it misses important vertex corrections that describe 
the effect of the electron
interactions on the coupling to the core hole potential $V({\bf r})$.  

In the extensively studied case of the 1D LL, for instance, crucial relevance 
of the vertex corrections can be seen
from the fact that although the faster-than-$1/t$ decay
of the fermion propagator $G({0}, t)\propto t^{-(2+g+g^{-1})/4}$, where $g$ 
is the Luttinger parameter ($0<g<1$ for short-range repulsive interactions),    
does imply a vanishing DOS, the FES retain 
their algebraic behavior with the $g$-dependent exponents \cite{3}. 

In the 1D case of purely forward scattering off the core hole potential, one can relate
the robustness of the FES to the fact that the density of particle-hole pairs 
with 
small 
momenta given by the integral of the density correlation function 
$\chi(\omega,q)$ maintains its non-interacting functional form: 
$\int^{Q<<k_F}_0 Im\chi(\omega, q)dq\propto \omega$ \cite{3}. 

This property is due to the known asymptotically exact cancellation between
the self-energy and vertex interaction corrections to the density correlation 
function at small momenta. 
In order to account for both types of corrections
 the previous studies \cite{3} resorted to the standard
method of 1D bosonization.

Furthermore, the bosonization approach allows one to treat an 
even more intricate 1D   
backward scattering problem. Unlike the case of pure forward 
scattering,
here the exponents controlling the power-law FES assume universal values 
which turn out to coincide with those corresponding to the unitary 
s-wave scattering $(\delta_0=\pi/2)$ in a Fermi liquid $(g=1)$ \cite{6}.   

Inspired by the success of 1D bosonization, there has been a strong recent 
effort
toward extending this method to higher dimensions \cite{7,8}. In what follows we 
employ the "tomographic"
version of this technique (see Refs.\cite{7}) which is capable of yielding 
asymptotically exact
results provided that fermions undergo predominantly forward scattering due to  
the interactions in the conduction band and the core hole potential.

Under these conditions the Lagrangian of the problem can be expressed solely
in terms of the bosonic partial 
densities $\rho_{\hat \Omega}^{\sigma}({\bf r},t)$ associated with different 
points on the Fermi surface
parameterized by the unit vector $\hat \Omega$. These degrees of freedom obey 
the infinite algebra
$$[\rho_{\hat \Omega}^{\sigma}({\bf r},t), \rho_{\hat 
\Omega^{\prime}}^{\sigma'}({\bf r}^{\prime},t)]=
K_{\sigma \sigma'}
\delta({\hat\Omega}-{\hat\Omega^{\prime}})({\bf \nabla}{\hat\Omega})\delta({\bf 
r}-{\bf r}^{\prime})
  \eqno(5)$$
Non-commutativity of the density operators at the same point of the Fermi 
surface
can be interpreted as the "chiral anomaly" which reflects, in compliance with 
the Luttinger theorem,
conservation of the total number of states enclosed by the nominal Fermi surface
in the presence of interactions between fermions. 

In the case of sufficiently weak spin-independent interactions
the $K$-matrix reduces to the unit: $K_{\sigma \sigma'}=\delta_{\sigma 
\sigma'}$. 
By introducing the $K$-matrix into the theory one 
can incorporate strong correlations between fermions of opposite spins
which modify the very kinematics of the system and can not be treated
as regular quadratic Landau-type terms in the effective Lagrangian.

Although the conduction band fermions $\psi_{\sigma}({\bf r}, t)$ 
couple to the forward scattering core hole potential only via the charge density 
operator

$\int d{\bf r}\sum_{\sigma}\psi^{\dagger}_{\sigma}({\bf r})V({\bf 
r})\psi_{\sigma}({\bf r})
\approx V\sum_{\hat\Omega}\sum_{\sigma}\rho_{\hat \Omega}^{\sigma}$, 
we keep the spin 
degrees of freedom as well,
thereby leaving open the possibility of incorporating the 
abovementioned exclusion principle-type spin-dependent correlations in the 
conduction band. 
Then the relevant part of the Lagrangian reads as 
$$
L= \sum_{\hat \Omega}\int d{\bf q} \rho_{\hat \Omega}^{\sigma}K^{-1}_{\sigma 
\sigma'}({\bf q}){i\partial_t
-v_F{\hat \Omega}{\bf q}\over {\hat \Omega}{\bf q}}\rho_{\hat 
\Omega}^{\sigma'}(-{\bf q}) - 
{1\over 2} \sum_{{\hat \Omega}, {\hat \Omega^\prime}}\int d{\bf q}
\rho_{\hat \Omega}^{\sigma}({\bf q})F_{{\hat \Omega},{\hat \Omega^\prime}}({\bf 
q})
\rho_{\hat \Omega^\prime}^{\sigma}( -{\bf q})+ $$
$$
+d^\dagger(i\partial_t-E_d) d - (V d^\dagger d -E_F)\sum_{\hat \Omega}\int d{\bf 
q}
\rho_{\hat \Omega}^{\sigma}({\bf q}) 
\eqno(6)$$
where we introduced a generalized (momentum and/or frequency-dependent)
Landau function $F_{{\hat \Omega},{\hat \Omega^\prime}}({\bf q})$
to describe the spin-independent ("residual") part of the forward scattering 
between the conduction band fermions. 

The quantity of primary physical interest is the X-ray absorption/emission 
intensity 
in the processes of electron transfer from the core level to the conduction band 
 and vice versa
$$
I(\omega)\propto Im \int^{\infty}_0dt e^{i(\omega+E_F-E_d) t} D(t), \ \
D(t-t^\prime)= <T d(t) \psi_{\sigma}({\bf 0}, t)\psi^{\dagger}_{\sigma}({\bf 0}, 
t^{\prime}) d^{\dagger}(t^{\prime})>
\eqno(7)$$
The amplitude (7) can be conveniently expressed as a gaussian 
average that has to be taken with respect to the Lagrangian (6) 
$$ 
D(t)\propto <\exp[ i\sum_{\hat \Omega}\int d{\bf q}\int d\omega
 (J_{\hat \Omega}(t,\omega, {\bf q})\phi_{\hat \Omega}^{\sigma}(-{\bf q})
+ V j(t,\omega)\rho_{\hat \Omega}^{\sigma}( -{\bf q})) ]> 
\eqno(8)$$
where the source densities 
$$J_{\hat \Omega}(t,\omega, {\bf q})=
{1-e^{i\omega t}\over \omega -v_F{\hat \Omega}{\bf q}}, \ \
j(t,\omega)={1-e^{i\omega t}\over \omega}
\eqno(9)$$ 
represent a conduction band fermion and an immobile
core hole, respectively, while $\phi_{\hat \Omega}({\bf q})$ is an auxiliary
Hubbard-Stratonovich variable conjugate to $\rho_{\hat \Omega}^{\sigma}({\bf 
q})$ 
$$
\int D \rho_{\hat \Omega}^{\sigma}
\exp[-{1\over 2}
%\sum_{{\hat \Omega}, {\hat \Omega^\prime}}\sum_{\sigma\sigma'} 
\rho_{\hat \Omega}^{\sigma}K^{-1}_{\sigma\sigma'}
{i\partial_t -v_F{\hat \Omega}{\bf q}\over {\hat \Omega}{\bf q}}\rho_{\hat 
\Omega}^{\sigma'}]=
\int D \phi_{\hat \Omega}^{\sigma}
\exp[-{1\over 2}
%\sum_{\hat \Omega} \sum_{\sigma\sigma'}
\phi_{\hat \Omega}^{\sigma}K_{\sigma\sigma'} {({\hat \Omega}{\bf q})\over
{i\partial_t- v_F({\hat \Omega}{\bf q})}}
\phi_{\hat \Omega}^{\sigma'} + 
%\sum_{\hat\Omega}\sum_{\sigma}
\phi_{\hat \Omega}^{\sigma}\rho_{\hat \Omega}^{\sigma}] 
\eqno(10)$$
In order to perform the averaging explicitly
one first has to compute the kernel of the angular decomposition for the density 
correlator 
$\chi(\omega,{\bf q})=
\sum_{{\hat \Omega},{\hat \Omega^\prime}}\sum_{\sigma\sigma'}
\chi_{{\hat \Omega},{\hat \Omega^\prime}}^{\sigma\sigma'}(\omega, {\bf q})$.
In practice, this amounts to inverting the operator  
$${\chi}^{-1}_{{\hat \Omega},{\hat \Omega^\prime}}(\omega,{\bf q})=
\delta({\hat \Omega}-{\hat \Omega^\prime}){\bf K}
{\omega -v_F({\hat \Omega}{\bf q})\over ({\hat \Omega}{\bf q})}
- {\bf 1}F_{{\hat \Omega},{\hat \Omega^\prime}}({\bf q})
\eqno(11)$$
Having carried out the gaussian functional averaging one arrives at the result
$$D(t)\propto \exp(-S(t)), \ \ S= S_{dos}+S_{exc}+S_{oc}   \eqno(12)$$ 
where the three different contributions to the total action $S(t)$ can be 
identified with, respectively, 
time dependence of the local conduction band propagator $G({\bf 0}, t)$
which is directly related to DOS: 
$$S_{dos}(t)= 
\int d\omega (1-\cos\omega t)
Im\int d{\bf q} {\chi_{{\hat \Omega},{\hat \Omega}}(\omega,{\bf q})
\over v^2_F({\bf q}{\hat \Omega})^2},   \eqno(13)$$
the "excitonic" term due to the interaction between the conduction band and
the core hole in the final state: 
$$S_{ex}(t)= V\int d\omega {1-\cos\omega t\over \omega} 
Im\sum_{\hat \Omega^\prime}\int d{\bf q}
{\chi_{{\hat \Omega},{\hat \Omega}^\prime}(\omega,{\bf q})
\over v_F({\bf q}{\hat \Omega})} 
\eqno(14)$$
and the OC term ${\it per} {\it se}$: 
$$S_{oc}(t)= V^2\int d\omega{1-\cos\omega t\over \omega^2} Im \sum_{{\hat 
\Omega},{\hat \Omega^\prime}}
 \int d{\bf q} \chi_{{\hat \Omega},{\hat \Omega^\prime}}(\omega,{\bf q})
\eqno(15)$$
We first comment on the use of Eqs.(13-15) in the free limit where  
one is supposed to recover the known results of Refs.\cite{2,4} by simply
putting the Landau function equal to zero.

Although for a sufficiently weak attractive core hole potential 
$(0<\delta_l<<1)$ 
we do reproduce the expected results for
 $S_{dos}=\log t$ and $S_{ex}=(2V\nu_F/\pi)\log t$, the expression for $S_{oc}$ 
 turns out to be formally divergent.
This non-physical divergence is solely due to our assumption of a strictly 
forward
scattering off the core hole, for in this extreme limit all 
the angular harmonics of the core hole
potential become equal: $V_l=V$, and therefore the sum   
$$\sum_{\hat \Omega}V^2=\sum_lV^2_l  \eqno(16)$$ 
 diverges. As a result, 
the vanishing OC factor $\exp(-S_{oc})$ in $I(\omega)$ outpowers any divergence 
that may  
result from the excitonic DOS enhancement at the location of the core hole,
and, as a result, all the FES are gone.
 
However, for any realistic core hole potential $V({\bf r})$ 
 the sum (16) is finite, and the overall energy dependence
of the X-ray absorption near the threshold assumes its classical form: 
$I(\omega) \propto \Theta(\omega)|\omega|^{-2\delta_0/\pi 
+2\sum_l\delta_l^2/\pi^2}$ \cite{2,4}.

It is worthwhile mentioning that the opposite case
of an isotropic core hole potential $(\delta_l=0$ for all $l\neq 0)$
allows one to construct an alternate bosonization scheme, thanks to the 
effectively 1D character of the radial 
motion of non-interacting fermions in the s-wave orbital channel \cite{9}.

Turning now to applications of the general formalism,
we first consider the limit of an exclusively "intra-tomograph"
spin-independent interaction that can only couple fermions with the same 
direction of the momenta:
${\bf K}={\bf 1}$ and $F_{{\hat \Omega}{\hat \Omega^{\prime}}}({\bf q})\propto  
\delta({\hat \Omega}-{\hat \Omega^{\prime}})$. 
A straightforward analysis shows that, similarly to the chiral 1D case, 
the only effect of such an interaction is a multiplicative renormalization of 
the Fermi velocity $v_F$ and, accordingly, the phase shifts: 
$\delta_l\rightarrow \delta_l/(1+F/v_F)$
that determine the, otherwise unaltered, FES exponents.

We note, in passing, that in order for a 
quasi-1D LL-like behavior to occur in $D>1$-dimensions, the spin-independent 
fermion interaction must couple each tomograph
$\hat \Omega$ to its antipodal one at $-\hat \Omega$:
$F_{{\hat \Omega}{\hat \Omega^{\prime}}}({\bf q})\propto 
\delta({\hat \Omega}-{\hat \Omega^{\prime}})+\delta({\hat \Omega}+{\hat 
\Omega^{\prime}})$.

In this situation the effective Luttinger parameter is given by the 
ratio
$g = v_F/c$ between the Fermi velocity and the speed of zero sound that can be 
read off from 
 the low-$q$ limit of the density correlator:
$ \chi(\omega, {\bf q})= \sum_{\hat\Omega}
({\hat \Omega}{\bf q})^2/(\omega^2 - c^2({\hat \Omega}{\bf q})^2 )$.

Again, in a close analogy with the situation in the 1D LL,
 coupling between the opposite points on the Fermi 
surface gives rise to, both, vanishing DOS $(\nu(\omega)\sim 
|\omega|^{(g+1/g-2)/4})$ and the modified, yet non-zero, FES exponents
$$P(\omega) \propto |\omega|^{ -1+2g^3\sum_l\delta_l^2/\pi^2}, 
\ \ I(\omega) \propto \nu(\omega)|\omega|^{- {2g\delta_0/\pi}+
2g^3\sum_l\delta_l^2/\pi^2} \eqno(17)$$
where the phase shifts are given by Eq.3 with a finite $\nu_F$ corresponding to 
the 
non-interacting case $g=1$. 

As follows from Eq.(17), 
repulsive interactions weaken the OC exponent via 
the reduced compressibility, thereby enhancing the overlap $<0|V>$ 
between the initial and the final ground states.

In principal, the asymmetrical profile of the photoemission peak $P(\omega)$ can 
be measured 
in absorption of hard X-rays that knock core electrons out of the system, 
whereas measuring $I(\omega)$ requires soft X-rays that can only promote
core electrons up to the conduction band. In practice, however, for the 
exponents in question
to be reliably deduced from a real X-ray spectrum, all the absorption lines and 
thresholds have to be well separated from one another. Apart from that,  one has 
to carry out 
a deconvolution of the measured intensity onto  
the FES factor and the extra 
factors describing recombination of the core hole as well as other 
mechanisms of the line broadening.

Furthermore, if the core hole potential can be approximated by the first two 
orbital harmonics: $V({\hat\Omega}-{\hat\Omega}^{\prime})\approx 
V_0+V_1Y_1({\hat \Omega}-{\hat\Omega}^{\prime})$,
then by measuring the two exponents in (17) and invoking the Friedel sum rule 
($2\sum_l\delta_l=\pi$ ) 
one might be able to extract the numerical values of, both, the relevant phase 
shifts $\delta_{0,1}$
and the Luttinger parameter $g$ that serves as a measure of the strength of
the LL-like correlations in the conduction band. 

Yet another evidence in favor of the above quasi-1D behavior would be 
the presence of the FES in the X-ray spectra even if the core hole 
has a finite, as opposed to infinite, mass.
Such an observation would be in a marked contrast with the  
situation in the $D>1$ Fermi liquids, where the hole recoil is known to reduce 
the number of
density excitations, thereby smearing the FES out \cite{9.1}.

Therefore, one indirect manifestation of the "tomographic" picture of fermion 
scattering
would be an observation of the FES in valence band photoemission
and a related algebraic decay (as opposed to the delta-function peak) of the 
spectral function of the  
mobile valence band hole \cite{9.2}.

An effective 1D LL-like behavior was conjectured in the very  
 beginning of the high $T_c$ saga
on the basis of such suggestive experimental findings about the normal state of 
the high $T_c$ cuprates as
the anomalous power-law decay of the optical conductivity $\sigma(\omega)\propto 
\omega^{-1+2\alpha}$ with $\alpha = 0.15\pm 0.05$,  the $\propto 1/T^2$ 
temperature dependence of the low-frequency
limit of the magnetic susceptibility at momentum ${\bf q}=(\pi,\pi)$, 
the $\propto\omega^{-1/2}$ tail of the electron spectral function $Im G({\bf k}, 
\omega)$, 
the linear c-axis tunneling density of states $dI/dV\sim V$, and others 
\cite{10}.

Thus far, traditional perturbative analyses did not turn in supporting evidence 
that
the above quasi-1D regime may indeed occur in 
a low-density weakly interacting system of fermions with a  
spherical Fermi surface. Nonetheless, 
one can expect the situation to be different
in paramagnetic lattice systems close to half-filling where the Hubbard-like 
no-double-occupancy constraint 
gives rise to the off-diagonal matrix elements
$K_{\sigma, -\sigma}=\phi/\pi$ in Eq.(5) which are 
proportional to the on-shell phase shift $-\pi/2\leq\phi\leq 0$ 
for a pair of fermions with opposite spins
\cite{7}. 

A simple analysis shows that even in the absence of any coupling between 
different "tomographs" the modified commutation relations (5) alone give rise to 
the 
anticipated spin-charge separation and change the OC 
exponent entering both    
$P(\omega)$ and $I(\omega)$ from $2\sum_l\delta^2_l/\pi^2$ to $(1+2\phi/\pi+{\sqrt 
{1+(2\phi/\pi)^2}})\sum_l\delta_l^2/\pi^2$ while leaving DOS unaltered
(of course, the latter gets affected, too, once coupling between different 
$\hat\Omega$'s is introduced). 
 
As a more traditional alternative to the kinematical constraint-type
interactions, it is believed that even a continuous (low-density) 
spinless fermion system may develop a NFL behavior
provided that the pairwise fermion interaction potential $U({\bf r})$ is 
sufficiently long-ranged.
 In terms of the Fourier transform $U({\bf q})=\lambda /q^{\eta}$ the sufficient 
condition 
reads as  $\eta\geq 2(D-1)$ \cite{11}. 
In Ref.\cite{11}, the 2D fermion distribution function 
was found to exhibit a non-analytical singularity instead of the finite step at 
the Fermi surface:
$\Delta n(k)\sim |k-k_F|^{\beta}$ with the exponent $\beta\propto 
\lambda^{1/2}$,  
which prompted speculations that all the other characteristics should exhibit 
a marked NFL behavior as well.

Surprisingly enough, our subsequent analysis of the physical 2D case reveals 
that  Fermi 
surface DOS may not vanish even in the presence
of the singular $U({\bf q})\sim 1/q^2$ interactions. Likewise, the FES remain 
largely intact.

We arrived at these conclusions by considering the situation when the Landau 
function
is independent of the location 
on the Fermi surface, although it may well be singular as a function of the 
transferred momentum. 
This, in turn, demands the solution of Eq.(11) to have the RPA form: 
$$F_{{\hat \Omega},{\hat \Omega^\prime}}({\bf q})=F({\bf q})={U({\bf q})\over 
1+U({\bf q})\Pi(\omega,{\bf q})},
\\\ \Pi(\omega, {\bf q})=\sum_{\hat\Omega}{({\hat \Omega}{\bf q})\over (\omega 
-v_F({\hat \Omega}{\bf q}))}
\eqno(18)$$
and, accordingly, the density correlator to be given by the RPA formula
$\chi=\Pi/(1+U\Pi)$.

Elaborating on Eq.(13) that represents the effect of the fermion interaction on 
DOS, we obtain
$$\nu(\omega)=
\nu_F \int_0^{\infty} {dt\over t} e^{i\omega t} \exp(- Im \int d\omega 
(1-\cos\omega t)
\int d{\bf q} {F(\omega,{\bf q})\over (\omega -v_F({\hat \Omega}{\bf q}))^2}) 
\eqno(19)$$
Our analysis of the momentum integral in Eq.(19) shows that the most important 
correction to DOS originates from the momenta
$q<<\omega/v_F$. In this regime the RPA Landau function is given by the 
expression 
$$F(\omega,{\bf q})= {\lambda\omega^2\over q^{\eta}\omega^2-\lambda q^{2}} 
\eqno(20)$$
where the pole reveals a sublinear plasmon dispersion $\omega\sim q^{1-\eta/2}$. 
 
After being plugged into Eq.(19), 
the Landau function (20) yields 

$S_{dos}\propto \beta\int_{1/t}^{E_F} 
d\omega\omega^{2(D-2)/(2-\eta)}$. 
Thus, for $\eta=2(D-1)$ and $1\leq D<2$ we obtain $S_{dos} \propto \log t$  
which does imply a power-law DOS:

$\nu(\omega)=\int^{\infty}_0 dt/t e^{i\omega t- S_{dos}}\propto 
|\omega|^{{\beta}/(2-D)}$ .

However, despite this suggestive behavior, in the physical dimension $D=2$ the 
plasmon mode develops a gap
and fails to generate any singular corrections to DOS. 
Nonetheless, this does not necessarily contradict the NFL-like behavior of the 
distribution 
function reported in Ref.\cite{11},
since a non-linear plasmon dispersion breaks the symmetry between energies and 
momenta.                                                                                       

Hence, in 2D the X-ray absorption intensity 
$I(\omega)\propto \nu_F\int_0^\infty e^{i\omega t -S_{oc+ex}}{dt/t}$ can only 
acquire a non-trivial energy dependence 
via the combined effect of the OC and "excitonic" terms in the total action (12) 
$$S_{oc+ex}= \int d\omega {1-\cos\omega t\over \omega^2} Im 
\sum_{\hat\Omega^{\prime}}
\int d{\bf q}(1+F(\omega,{\bf q})\Pi(\omega, {\bf q}))$$
$$ 
(2\delta_0{\omega\over ({\bf q}{\hat\Omega})}\delta({\hat \Omega}-{\hat 
\Omega^{\prime}}) -\sum_l\delta^2_l)
{({\bf q}{\hat\Omega^{\prime}})\over {\omega -v_F({\bf 
q}{\hat\Omega^{\prime}})}}$$
$$
=(2\delta_0-\sum_l\delta^2_l) \int {d\omega\over \omega}(1-\cos \omega t)
\int {d^{d-1}{\bf q}_{\perp}\over 1+\lambda/q^{\eta}}
\eqno(21)$$
which, in contrast to the momentum integral in (19), receives its main 
contribution from the momenta $q>\omega/v_F$  
while the contribution of the momenta $q<\omega/v_F$ can only become important 
for a  
1D short-range interaction $(\eta=0)$.

At this point, a word of caution is in order. 
After having obtained Eq.(21), one may question the use of bosonization employed 
for 
its derivation.
Indeed, unlike Eq.(19), the region of momenta contributing to the integral over 
${\bf q}_{\perp}$ does not scale down to zero 
in the asymptotic limit $\omega\rightarrow 0$. In fact, one has to set the 
integration limit to be of order the inverse Fermi surface curvature
for the classical results of Refs.\cite{2,4} to be recovered for $\lambda =0$.

Nonetheless, we believe that, being formulated solely in terms of the 
density correlator (or, more precisely, its angular decomposition 
$\chi_{{\hat \Omega},{\hat \Omega}'}(\omega, {\bf q}) )$, the general 
Eqs.(13-15) 
remain qualitatively valid, thereby indicating that, although the singular 
repulsive 
interactions in the conduction band 
do affect the FES exponents via the reduced compressibility, they do not 
eliminate the FES altogether.

In a pursuit of a genuine $D>1$ NFL behavior, one can proceed further by  
incorporating  a finite  inelastic scattering time $\tau(\omega)$ into the 
theory.
The latter is proportional 
to the optical conductivity $\sigma(\omega)$ at frequencies $\omega>>1/\tau(T)$,  
while in the opposite limit it saturates at a finite value $\tau(T)$.

As a result, the density modes acquire  
an overdamped character even in the absence of any real disorder (elastic 
scattering), and the pole of the density correlator 
$\chi(\omega, {\bf q})={\bf q}^2/(\omega^2-c^2{\bf q}^2+i\omega/\tau(\omega)) $ 
moves to the imaginary axis. 

In order to gain a better insight into the overdamped regime, we consider a 
generic Landau function
$F(\omega, {\bf q})=q^m/(i\omega+q^n)$ whose functional form is inspired by the 
studies of the compressible Quantum Hall
states at even-denominator filling fractions \cite{13} as well as 
quantum critical points in itinerant magnets and Mott insulators \cite{14}. 

Computing the integrals (13-15)
we obtain the following asymptotics for the interaction terms that 
add to the bare $\log t$ terms corresponding to the non-interacting limit
$$ \Delta S_{dos}\propto t^{(3-n-m-D)/n},\ \ \Delta S_{oc}\propto 
t^{2-(m+D)/n},\ \
 \Delta S_{ex}\propto t^{1-(D-1+m)/n} \eqno(22)$$
For $n=1$ all of the above long time asymptotics coincide and, therefore,
all the three terms in (22) are equally important,
whereas for $n>1$ it is the OC term that 
dominates over the other two. Moreover, for 
$D+m<2n$ the divergence of $\Delta S_{oc}$ is strong 
enough for the FES to be replaced by a rounded edge whose profile appears to be 
strongly non-analytical if the parameters 
satisfy the additional condition $n<D+m$: 
$$I(\omega)\propto \exp(-|\omega|^{(-2n+D+m)/(D+m-n)}) \eqno(23)$$
The non-algebraic behavior described by
Eq.(23) gives the final result as long as the core hole potential can be 
considered as a predominantly forward scatterer, and the bosonized theory (6) remains 
gaussian. The same reservation applies to the conduction band interactions that 
we chose to be quadratic in terms of the bosonic density modes.

For a realistic core-hole potential, however, the non-forward-scattering 
contribution 
is likely to become increasingly more and more important 
at $\omega \rightarrow 0$. It remains to be seen whether or not an algebraic  
behavior controlled by some universal
exponent gets restored at the lowest energies (see Ref.\cite{6} for the analysis
of the problem of 1D backscattering where Eq.(15) yields 
$S_{oc}\propto t^{2(1-g)}$). 

Indeed, in the framework of the conventional cluster expansion
that does not rely on
any bosonic representation, 
Eqs.(15) appears to be merely the first significant
$(k=2)$ term in the infinite series $S_{oc}=\sum_kS^{(k)}_{oc}$,
where $S^{(k)}_{oc}\propto V^k$ is given by a convolution of the $k$ functions 
$\chi(\omega, {\bf q})$.

In analogy with  the 1D backscattering problem, one may expect  
that a universal asymptotic behavior $S_{oc}\propto \log t$ with a prefactor 
independent of
the strength of a generic (non-forward-scattering) core-hole potential occurs   
in the case when 
each $k^{th}$ order term of the cluster expansion (unless it vanishes) has 
a stronger divergence than the $(k-1)^{th}$ one at $t\rightarrow \infty$.
However, regardless of whether or not the photoemission line broadening  
washes this regime out, the universal exponents themselves do not
provide much insight into a nature of the electron correlations in the 
conduction band. 

Putting it another way, for a generic core-hole
potential a simplified picture of 
 the conduction band (suggested by the lowest order of the cluster expansion 
Eq.(15))
  as a "Fermi bath" characterized solely by
its spectral density of particle-hole excitations 
$A(\omega)=\int d{\bf q}Im\chi(\omega,{\bf q})$
can only hold at some intermediate time scales, until the 
 higher terms $S^{(k)}$ with $k>2$ come into play.    

Before concluding, we briefly comment on the effect of a finite temperature that 
tends to further amplify the divergence of the expressions given by Eqs.(13-15)  
via the occupation number
of thermally excited collective modes $N(\omega)\approx T/\omega$. 
However, the resulting hardening of the OC
contribution may not be detectable, since it will be superimposed with the 
smearing of the FES features
stemming from 
 the free fermion Green functions $G({\bf 0}, t)\propto 1/t \rightarrow \pi 
T/\sinh(\pi Tt)$.

To summarize, in this paper we  
considered the problem of the X-ray absorption in the presence of 
 interactions in the conduction band. By restricting our attention to the case 
of the forward 
scattering core hole potential
we were able to incorporate the fermion interactions in the framework of the 
bosonization approach.
We illustrated the general method with such examples as the $D>1$-dimensional 
"tomographic" LL, fermions  with a singular pairwise potential, and coupling via 
an overdamped collective 
mode. 
We found that, in general, 
neither a vanishing single-particle DOS precludes the existence of the FES, nor 
can one 
guarantee that the FES remain algebraic at all energy scales.
Altogether, these findings testify in favor of the FES features in the X-ray 
spectra offering an independent 
way to probe many-body correlations in strongly
interacting Fermi systems.

The authors acknowledge the support from the NSF under the Grant DMR-9104873
and hospitality at the Aspen Center for Physics where this work was completed.


\begin{references}

\bibitem{1} M.R.Norman {\it et al}, Nature {\bf 392}, 157 (1998); Phys.Rev.{\bf 
B57}, R11093 (1998).
\bibitem{2} P.W.Anderson, Phys.Rev.Lett.{\bf 18}, 1049 (1967).
\bibitem{4} G.D.Mahan,  Phys.Rev.{\bf 163}, 612 (1967).
\bibitem{3} T.Ogawa {\it et al}, Phys.Rev.Lett.{\bf 68}, 3638 (1992); D.Lee and 
Y.Chen, ibid {\bf 69}, 1399 (1992).
\bibitem{6} A.Gogolin, Phys.Rev.Lett.{\bf 71}, 2995 (1993); N.V.Prokofiev, 
Phys.Rev.{\bf B49}, 2148 (1994);
C.L.Kane {\it et al}, ibid {\bf B49}, 2253 (1994).
\bibitem{5} P.Nozieres and C.T.De Dominicis, Phys.Rev.{\bf 178}, 1097 (1969).
\bibitem{7} A.Luther, Phys.Rev.{\bf B19}, 320 (1979); P.W.Anderson, 
Phys.Rev.Lett.{\bf 64}, 1839 (1990); 
ibid {\bf 68}, 2306 (1991); 
F.D.M.Haldane, Helv.Phys.Acta {\bf 65}, 152 (1992); P.W.Anderson and 
D.V.Khveshchenko, Phys.Rev.{\bf B52}, 16540 (1995).
\bibitem{8} A.Castro-Neto and E.Fradkin, Phys.Rev.{\bf B49}, 10877 (1994);
A.Houghton, H.J.Kwon, and J.B.Marston, ibid {\bf B50}, 1351 (1994);
P.Kopietz and K.Schonhammer, ibid {\bf B52}, 10877 (1995). 
\bibitem{9} K.D.Schotte and U.Schotte, Phys.Rev.{\bf 182}, 479 (1969).
\bibitem{9.1} J.Gavoret {\it et al}, J. de Phys. (Paris) {\bf 30}, 987 (1969). 
\bibitem{9.2} T.Kopp, A.E.Ruckenstein, and S.Schmitt-Rink, Phys.Rev.{\bf B42}, 
6850 (1990). 
\bibitem{10} P.W.Anderson, {\it The Theory of Superconductivity in High $T_c$ 
Cuprates} 
(Princeton University Press, Princeton, 1997). 
\bibitem{11} P.Bares and X.G.Wen, Phys.Rev.{\bf B48}, 8636 (1993).
\bibitem{13} B.I.Halperin, P.A.Lee, and N.Read, Phys.Rev.{\bf B47}, 7312 (1993).
\bibitem{14} J.A.Hertz, Phys.Rev.{\bf B14}, 1165 (1976).


\end{references}
\end{document}